\documentclass{article}
\title{A Computation of The Mass Spectrum of Mesons and Baryons }

\author{Sze Kui Ng\\
{\small Department of Mathematics,
Hong Kong Baptist University, Hong Kong}
}

\begin{document}
\date{}

\maketitle
\begin{abstract}
In this paper we give a computation of the mass spectrum of mesons and baryons. By this computation we show that there is
a consecutive numbering of the mass spectrum of mesons and baryons. We show that in this numbering many stable mesons 
and baryons are assigned with a prime number. 


\end{abstract}

\section{Introduction}\label{sec00}

In this paper we give a computation of the mass spectrum of mesons and baryons. We show that there is
a consecutive numbering of the mass spectrum of mesons and baryons. Also we show that in this numbering many stable mesons 
and baryons are assigned with a prime number. 

We first consider the mesons listed in the Particle Data Group (PDA)\cite{PDA}. Let us start
with the $\pi$ meson. Let us choose $45$ as a rate. Then we have $3 \times 45=135$ which gives
the mass $135 Mev$ of the $\pi^0$ meson. We notice that $3$ is a prime number and we assigned it to the  $\pi^0$ meson. Then we consider the koan mesons. We have that $11 \times 45= 495$. This approximates well the experimental mass $493.7 Mev$ of the $K^+ $ and the $K^-$ mesons. We notice that $11$ is a prime number and we assign it to the koan mesons. Let us then consider the $\eta$ meson. We have $13\times 45= 585$ which approximates quite well the experimental mass $548.8 Mev$ of the 
$\eta$ meson. Here let us make a more precise estimate of the mass of $\eta$ meson. Let $45$ be deviated
by $3$ to $42$. This deviation may be considered as due to the effect of spin and orbital angular momentum for mass splitting. Then we have $13\times 42=546$ which approximates well the experimental mass $548.8 Mev$ of the $\eta$ meson. We have that $13$ is a prime number and we assign it to the $\eta$ meson. This then gives the mass spectrum of the first Eightfold Way of mesons. We shall later consider the $\eta^{\prime}$ meson of the first nonet.

Let us then consider the next octet of vector mesons. Let us first consider the $\rho$ meson. We have $17\times 45= 765$. This approximates the experimental mass $770 Mev$ of the $\rho^0$ meson. Let us then consider the $\omega$ meson. Let $45$ be deviated by $1$ to $46$. As above this deviation may be considered as due to the effect of spin and orbital angular momentum for mass splitting. Then we have $17\times 46 = 782$. This approximates well the experimental mass $783 Mev$ of the $\omega$ meson. We notice that both the $\rho^0$ meson and the $\omega$ meson correspond to the same prime number $17$. Let us then consider the $K^*(892)$ meson. We have $19 \times 45=852$. This approximates  quite well the experimental mass $892 Mev$ of the $K^*(892)$ meson. Here let us make a more precise estimate of the $K^*$ mesons. Let $45$ be deviated by $2$ to $47$. Then we have $19 \times 47=890$. This approximates well the experimental mass $892 Mev$ of the $K^*(892)$ meson. we have that the $K^*(892)$ meson correspond to the prime number $19$. Let us then consider the $\phi$ meson. We have $23\times 45= 1035$. This approximates well the experimental mass $1020 Mev$ of the $\phi$ meson. We notice that $23$ is a prime number and we assign it to the $\phi$ meson.
This thus gives the mass spectrum of the nonet of vector mesons. 

Let us then consider the $\eta^{\prime}$ meson of the first nonet. Let $45$ be deviated by $3$ to $42$.
Then we have $23\times 42=966$. This approximates well the experimental mass $958 Mev$ of the $\eta^{\prime}$ meson. This thus gives the mass spectrum of the first nonet of mesons.



We notice that these two nonets are finally numbered at the prime number $23$. This is interesting because the next prime number is $29$ which is relatively far apart from the prime number $23$. We also notice that the consecutive prime numbers $11, 13, 17, 19$ and $23$ have all been assigned to the mesons of the first two nonets of mesons which are stable mesons. 


The $\phi$ meson is considered as a meson of the form $s\overline{s}$ where $s$ denotes the strange quark. 
Then it is interesting to note that both the mesons $\phi$ and $\eta^{'}$ correspond to the prime number $23$.
This shows that there is a relation between these two mesons. This agrees with the usual quark content of these two mesons that these two mesons are considered to have a component of $s\overline{s}$. Later we show that the mesons $a_0(980)$ and $f_0(980)$ also correspond to the prime number $23$. From this we may conclude that these two mesons $a_0(980)$ and $f_0(980)$ should also relate to $s\overline{s}$. This agrees with the experiments that these two mesons lie very close to the opening of the $K\overline{K}$ channel \cite{Abe}-\cite{Kam}.

Let us then considered the $J/ \psi$ meson $c\overline{c}$ for the charm quark $c$. 
We have $67\times 46=3082$. This approximetes quite well the experimental mass $3096 Mev$ of
the $J/ \psi$ meson. Thus $J/ \psi$ is assigned with the prime number $67$.

Let us consider more on the $J/ \psi$ meson.
For the 
$J/ \psi$ meson let us choose $23\times 45$ as another rate. Then we have $3\times 23\times 45= 3105$. This approximates well the experimental mass $3096 Mev$ of the $J/ \psi$ meson. Thus the prime number $3$ also corresponds to the $J/ \psi$ meson which is a stable meson. We notice that
$23\times 45 Mev =1035 Mev$ is just the mass for the $\phi$ meson. This gives a relation between the strange quark and the charm quark.

Let us then consider the $\Upsilon$ meson which is of the form $b\overline{b}$ where $b$ denotes the beauty (or bottom) quark. We have $211 \times 45=9495$. This approximates quite well the experimental mass $9460 Mev$ of the $\Upsilon$ meson. We also notice that $211$ is a prime number. Then let us consider the $\Upsilon(10023)$ meson which is the first excite state of the 
$\Upsilon$ meson. We have $223 \times 45=10035$. This approximates well the experimental mass $10023 Mev$ of the $\Upsilon^{\prime}$ meson. It is interesting to notice that $223$ is also a prime number and that it is the prime number next to the prime number $211$ for the $\Upsilon$ meson.

Let us consider two more excited states $\Upsilon(10350)$ and $\Upsilon(10570)$ of the $\Upsilon$ meson. We have $230\time 45=10350$ and $235\times 45=10575$. This approximete well the experimental masses of $\Upsilon(10350)$ and $\Upsilon(10570)$ respectively. However we notice that $230$ and $235$ are not prime numbers.

Thus we see that all the basic mesons except the meson $t\overline{t}$ for the top quark $t$ correspond to a prime number. For the meson $t\overline{t}$ because its mass  is very large that we need a more accurate measurement of the value of its mass to determine the prime number for this meson $t\overline{t}$. We have $7993\times 45=359685$ which approximately gives $360 Gev$ of the experimental mass of $t\overline{t}$. Then the prime number $7993$ could be approximately for $t\overline{t}$.

Continuing in this way with $45$ or its deviations as a rate we have that all the experimental masses of mesons in the table of mesons can well be approximated  by this computation. For simplicity we list the computational result in a form of table. In this table we list together the light mesons and strange mesons which are separately listed in the Particle Data Group (PDA) \cite{PDA}.  In this table in the column of computed mass the first number of the product is as a consecutive number while the second number is as a deviation from the number $45$. In this table the prime consecutive numbers are in bold face.
From this table we see that there are more mesons which  could be corresponded to a prime consecutive number. This is an evidence that prime numbers give more stable mesons than nonprime numbers. Also it is interesting to notice that the consecutive numbers from $26$ to $53$ have all been assigned to the mesons.

In this table of mesons  for the mesons $a_0(980)$ and $f_0(980)$ we have $23\times 43=989$.
This approximates quite well the experimental mass $980 Mev$ of $a_0(980)$ and $f_0(980)$. Thus the prime number $23$ is also assigned to $a_0(980)$ and $f_0(980)$. By the same reason as for the $\eta^{'}$ meson we have that $a_0(980)$ and $f_0(980)$ are related to $\phi$ and their quark contents sould have a component of $s\overline{s}$. This agrees with the experiments of these two mesons \cite{Abe}-\cite{Kam}.

\begin{displaymath}
\begin{array}{|c|l|} \hline
\mbox{Meson}& \mbox{Computed mass} 
\\ \hline
\pi(135) & {\bf 3}\times 45=135 \\ \hline

K(494) & {\bf 11}\times 45=495  
\\ \hline

\eta(549) &{\bf 13}\times 42=546 \\ \hline
\rho(770) &{\bf 17} \times 45=765 \\ \hline
\omega(783) &{\bf 17}\times 46=783  
\\ \hline

K^{*}(892) &{\bf 19}\times 47=893  \\ \hline
\eta^{\prime}(958) &{\bf  23}\times 42=966   \\ \hline
 a_0(980)& {\bf 23}\times 43=989 \\ \hline
 f_0(980) &{\bf 23}\times 43=989 \\ \hline

\phi(1020) & {\bf 23}\times 45=1035  
\\ \hline

h_1(1170) & 26\times 45=1170 
\\ \hline

b_1(1235) & 27 \times 45=1242 
\\ \hline

a_1(1260) & 28 \times 45=1260 \\ \hline

f_1(1285) & 28 \times 46=1288 
\\ \hline

a_2(1320)& 28\times 47=1316 \\ \hline

f_2(1270) & {\bf 29} \times 44= 1276
\\ \hline

\pi(1300) & {\bf 29} \times 45=1305 \\ \hline

 f_0(1370)&{\bf 29} \times 47=1363 \quad({\bf 31}\times 44=1364) \\ \hline

 f_1(1420)& {\bf 29}\times 49=1421 \quad( {\bf 31}\times 46=1426)\\ \hline

 h_1(1380)& 30\times 46= 1380 \\ \hline

K_1(1270) & {\bf 31 }\times 41=1271\\ \hline

\eta(1295) & {\bf 31}\times 42=1302  
\\ \hline

K_1(1400)& {\bf 31}\times 45=1395 \\ \hline

 f_2(1430)&{\bf 31}\times 46= 1426
\\ \hline

 K_0^*(1430)&{\bf 31}\times 46=1426   
\\ \hline

 K_2^*(1430)& {\bf 31}\times 46=1426 \\ \hline

 K(1460)& {\bf 31}\times 47= 1457  
\\ \hline

\pi_1(1405)&32\times 44=1408 \quad({\bf 31}\times 45=1395)\\ \hline

 f_1(1510)& 32\times 47=1504 \quad( 33\times 46=1518) \\ \hline

 \omega(1420)& 33\times 43=1419 \quad({\bf 31}\times 46=1426, {\bf 29}\times 
49=1421)\\ \hline

 \rho(1450)& 33\times 44=1452 \quad({\bf 31}\times 47=1457) \\ \hline

 a_0(1450)& 33\times 44=1452 \\ \hline

 f_0(1500)& 33\times 45=1495   
\\ \hline

 f_2(1565)& 34\times 46=1564 \quad({\bf 37}\times 42= 1554)\\ \hline

 K_2(1580)& 35\times 45=1575 \\ \hline

 \eta_2(1645)& 35\times 47=1645 \\ \hline

 X(1650)& 36\times 46=1656 \\ \hline

K_1(1650)& 36\times 46=1656 \\ \hline

 \rho_3(1690) & 36\times 47=1692\\ \hline

K_2(1770) & 36\times 49=1764\\ \hline
\end{array}
\end{displaymath}

\begin{displaymath}
\begin{array}{|c|l|} \hline
\mbox{Meson}& \mbox{Computed mass} 
\\ \hline

 K^*(1410)& {\bf 37}\times 38=1406 \quad(30\times 47=1410) \\ \hline

 \eta(1440)& {\bf 37}\times 39=1443 \quad( 32\times 45=1440) \\ \hline

 f_2^{'}(1525)& {\bf 37}\times 41=1517 \quad(34\times 45=1530) \\ \hline

 X(1600)& {\bf 37}\times 43=1591   
\\ \hline

 \omega(1600)& {\bf 37}\times 43=1591   
\\ \hline

 \omega_3(1670)& {\bf 37}\times 45=1665 \\ \hline

 \pi_2(1670)& {\bf 37}\times 45=1665 \\ \hline

\rho(1700) &  {\bf 37}\times 46=1702 \\ \hline
X(1775) & {\bf 37}\times 48=1776 \\ \hline

f_0(1710) & 38\times 45=1710 \\ \hline

K_3^*(1780) &38\times 47=1786 
\\ \hline

K_2(1820) & 38\times 48=1824\\ \hline

\eta_2(1870)&39\times 48=1872
\\ \hline

X(1910) & 39\times 49=1911 
\\ \hline

X(2000)
 & 40\times 50=2000  
\\ \hline

 a_4(2040)& 40\times 51=2040  
\\ \hline

 \phi(1680)& {\bf 41}\times 41=1681 \quad(42\times 40 =1680)\\ \hline

 K^*(1680)& {\bf 41}\times 41=1681 \quad(42\times 40 =1680)\\ \hline

 f_2(1640)& {\bf 41}\times 40 =1640 \quad({\bf 37}\times 44=1628,38\times 43=1634) \\ \hline

f_2(1810)&{\bf 41}\times 44=1804 \\ \hline

\phi_3(1850)
& {\bf 41}\times 45=1845 \\ \hline

K_0^*(1950)&{\bf 41}\times 48=1968 \quad( {\bf  43}\times 45=1935) \\ \hline

f_2(1950) &{\bf 41}\times 48=1968 \quad({\bf 43}\times 45=1935) \\ \hline

 f_2(2010)
& {\bf 41}\times 49= 2009  
\\ \hline

K_4^*(2045)
& {\bf 41}\times 50=2050  
\\ \hline

f_4(2050)
& 42\times 49=2058  
\\ \hline

\eta(1760)& {\bf 43}\times 41=1763 \quad( 40\times 44=1760)  \\ \hline

\pi(1800) &{\bf 43}\times 42=1806 \quad( 40\times 45=1800) \\ \hline

 f_0(2020)& {\bf 43}\times 47=2021  
\\ \hline

 f_0(2060)
& {\bf 43}\times 48=2064  
\\ \hline

\pi_2(2100)
& {\bf 43}\times 49=2107  
\\ \hline

f_2(2150)
& {\bf 43}\times 50=2150  
\\ \hline

\rho(2150)
& {\bf 43}\times 50=2150  
\\ \hline

K_2^*(1980) &44\times 45=1980  \quad({\bf 43}\times 46=1978, 46\times 43=1978) 
\\ \hline

 \rho_3(2250)
& 45\times 50=2250  
\\ \hline

f_2(2300)
& 46\times 50 =2300  
\\ \hline

 f_4(2300)
&  46\times 50 =2300  
\\ \hline

K(1830)& {\bf 47}\times 39=1833 \quad( 39\times 47=1833) \\ \hline

 f_J(2220)
&{\bf 47}\times 47=2209  
\\ \hline

K_2(2250)
& {\bf 47}\times 48=2256 \quad( 50\times 45=2250)  
\\ \hline

 \rho_5(2350)& {\bf 47}\times 50=2350  
\\ \hline

K_4(2500)
& 48\times 52=2496 \quad({\bf 47}\times 53=2491, 50\times 50= 2500)  
\\ \hline

a_6(2450)
&  49\times 50=2450
\\ \hline

f_0(2200)
& 50\times 44=2200  
\\ \hline

f_2(2300)
& 51\times 45 =2295  
\\ \hline

f_2(2340)
&52\times 45=2340  
\\ \hline

\eta(2225)
&{\bf 53}\times 42= 2226  
\\ \hline

K_5^*(2380)
&{\bf  53}\times 45=2385 
\\ \hline

K_3(2320)& 54\times 43=2322  
\\ \hline

 f_6(2510)& 57\times 44=2508  
\\ \hline

K(3100)
&{\bf 67}\times 46=3082  
\\ \hline

X(3250)
&{\bf 71}\times 46=3266  
\\ \hline
\end{array}
\end{displaymath}

Let us then consider the mass spectrum of baryons. In contrast to the rate $45$ for mesons let us choose $72$ as the rate for baryons.

 Let us start from the the neutron $n=N(939)$. We have $13\times 72=936$. This approximates well the experimental mass $939 Mev$ of the neutron $n=N(939)$. We have that the prime number $13$ is assigened to proton and neutron.
Then we consider $\Lambda (1115)$. Let $72$ be deviated to $66$. Then we have $17\times 66 =1122$. This approximates well the experimental mass $1115 Mev$ of $\Lambda (1115)$. We have that the prime number $17$ is assigned to $\Lambda (1115)$.  Then we consider $\Sigma(1193)$. We have $17\times 70=1190$ where $70$ is as a deviation of $72$. This approximates well the experimental mass $1193 Mev$ of $\Sigma(1193)$. Thus the prime number $17$ is also assigned to $\Sigma(1193)$. 
Then we consider $\Xi(1317)$. We have $19\times 69=1311$ where $69$ is as a deviation of $72$. 
This approximates well the experimental mass $1317 Mev$ of $\Xi(1317)$. The prime number $19$ is assigned to $\Xi(1317)$. This gives  the mass spectrum of the first Eightfold Way of baryons.

Let us then consider the first decuplet of baryons. Consider first $\Delta(1232)$. We have $17\times 72=1224$. This approximates well the experimental mass $1232 Mev$ of $\Delta(1232)$. The prime number $17$ is assigned to $\Delta(1232)$. Then we consider $\Sigma(1385)$. We have $19\times 73=1387$ where $73$ is as a deviation of 72. This approximates well the experimental mass $1385 Mev$ of $\Sigma(1385)$. The prime number $19$ is then assigned to $\Sigma(1385)$. Then we consider $\Xi(1530)$. 
Let $72$ be deviated by $1$ to $73$. Then we have $21\times 73=1533$. This approximates well the experimental mass $1530 Mev$ of $\Xi(1530)$. The number $21$ is assigned to $\Xi(1530)$. 
(On the other hand we also have $19\times 81=1539$ which also approximates 
$1530$ well. Thus the prime number $19$ may also be assigned to $\Xi(1530)$.
However $81$ is a larger deviation from $72$ than $73$).
Finally we consider $\Omega(1672)$. Let $72$ be deviated by $1$ to $73$. Then we have $73\times 23=1679$. This approximates well the experimental mass $1672 Mev$ of $\Omega(1672)$. Thus the prime number $23$ is assigned to $\Omega(1672)$. This gives  the mass spectrum of the first decuplet of baryons.

It is interesting to notice that as the $\phi$ meson we have that the $\Omega(1672)$ baryon  also corresponds to the prime number $23$. Also the  second octet of vector mesons and the  first decuplet of baryons are both assigned with prime numbers $17, 19$ and $23$.  This shows that the  second octet of mesons corresponds to the first decuplet of baryons.

Continuing in this way we have that all the experimental masses of baryons in the table of baryons in the Particle Data Group (PDA) can well be approximated  by this computation. For simplicity we list the computational result in a form of table.  In this table in the column of computed mass the first number of the product is as a consecutive number. Also the prime consecutive numbers are in bold face.


\begin{displaymath}
\begin{array}{|c|c|c|c|} \hline
 N \mbox{baryon}& \mbox{Computed mass} 
 & \Delta \mbox{baryon}& \mbox{Computed mass}
\\ \hline

 N(939)& {\bf 13}\times 72= 936 & \Delta(1232)  & {\bf 17}\times 72=1224 \\ \hline

 N(1440)& {\bf 19}\times 76=1444 & \Delta(1550) & {\bf 21}\times 74=1554 \\ 
 & 20\times 72=1440&  &  \\ \hline

 N(1520)& {\bf 19}\times 80=1520& \Delta(1600) & 22\times 73=1606  \\ 
\hline

 N(1535)& {\bf 19}\times 81=1539 & \Delta(1620) & {\bf 23}\times 70=1610 \\ \hline

 N(1540)& {\bf 23}\times 67=1541 & \Delta(1700) & {\bf 23}\times 74=1702 \\
 &22\times 75=1540 & &25\times 68=1700 \\ \hline

 N(1650)& {\bf 23}\times 72=1656 &\Delta(1905) &{\bf 23}\times 83=1909 \\ 
 &22\times 75=1650 & & \\ \hline

 N(1675)& {\bf 23}\times 73=1679&\Delta(1930) &{\bf 23}\times 84=1932 \\ 
 &25\times 67=1675 & & \\ \hline 

 N(1680)& {\bf 23}\times 73=1679&\Delta(1920) & 24\times 80=1920 \\ 
 &24\times 70=1680& & \\ \hline

 N(1700)& {\bf 23}\times 74=1702 &\Delta(1940) & 24\times 81= 1944\\ 
 &25\times 68=1700 & & \\ \hline 

 N(1710)& {\bf 23}\times 74=1702 &\Delta(1900) & 25\times 76=1900 \\ 
 &24\times 71=1704 & & \\ \hline 

 N(1720)&{\bf 23}\times 75=1727 &\Delta(1910) &{\bf 29}\times 66=1914 \\ \hline 

N(1900)&25\times 76=1900& \Delta(1950) &{\bf 29}\times 67= 1943\\ \hline

 N(1990)& 28\times 71= 1988& \Delta(2150) &{\bf 29}\times 74= 2146\\ \hline

 N(2000)&{\bf 29}\times 69= 2001 & \Delta(2160) & 30\times 72= 2160\\ \hline

 N(2080)&{\bf 29}\times 72= 2088 & \Delta(2200) &{\bf 31}\times 71= 2201\\ \hline

 N(2100)& 30\times 70= 2100 & \Delta(2300) &{\bf 31}\times 74= 2294 \\ \hline

 N(2190)&{\bf 29}\times 75= 2175& \Delta(2350) &{\bf 31}\times 76= 2356 \\ 
        & 30\times 73= 2190&   &  \\  \hline

 N(2200)&{\bf 31}\times 71= 2201& \Delta(2400) & 30\times 80=2400 \\ \hline

 N(2220)&{\bf 31}\times 72= 2232& \Delta(2420) & {\bf 31}\times 78= 2418 \\ \hline

 N(2250)&{\bf 31}\times 73=2263 & \Delta(2500) &{\bf 31}\times 81=2511 \\ \hline

 N(2600)&{\bf 37}\times 70=2590 &\Delta(2850) &{\bf 37}\times 77= 2849 \\ \hline

 N(2700)&{\bf 37}\times 73=2701 &\Delta(2750) &{\bf 41}\times 67=2747  \\ \hline

 N(2800)&{\bf 37}\times 76=2812 & \Delta(2950) &{\bf 41}\times 72= 2952 \\ \hline

 N(3000)&{\bf 41}\times 73=2993 &  \Delta(3230) &{\bf 43}\times 75= 3225 \\
\hline

 N(3030)&{\bf 41}\times 74=3034 &  \\ 
        &{\bf 43}\times 71=3053 &   &  \\ \hline

 N(3245)&{\bf 47}\times 69=3243 &  &  \\ \hline

 N(3690)&{\bf 53}\times 70=3710 &  &  \\ \hline

 N(3755)&{\bf 53}\times 71=3763 &  &  \\ \hline
\end{array}
\end{displaymath}

\begin{displaymath}
\begin{array}{|c|c|c|c|} \hline
\Lambda \mbox{baryon}& \mbox{Computed mass}
 & \Sigma \mbox{baryon}& \mbox{Computed mass} 
\\ \hline
 \Lambda(1115)&{\bf 17}\times 66=1122 & \Sigma(1193)&{\bf 17}\times 70=1190\\ \hline

 \Lambda(1405) &{\bf 19}\times 74=1406& \Sigma(1385) &{\bf 19}\times 73=1387\\ \hline

 \Lambda(1520) &{\bf 19}\times 80=1520& \Sigma(1480) &{\bf 23}\times 64=1472\\ \hline

 \Lambda(1600) &{\bf 23}\times 70=1610& \Sigma(1560) &{\bf 23}\times 68=1564\\ 
 &22\times 73=1606 & & \\ \hline

 \Lambda(1670) &{\bf 23}\times 73=1679& \Sigma(1580) &{\bf 23}\times 69=1587\\ 
 &22\times 76=1672  & & \\ \hline

 \Lambda(1690) &{\bf 23}\times 74=1702& \Sigma(1620) &{\bf 23}\times 70=1610\\ 
 &22\times 77=1694 & & \\ \hline

 \Lambda(1800) &{\bf 23}\times 78=1794& \Sigma(1660) &{\bf 23}\times 72=1656\\ 
 &24\times 75=1800 & & \\ \hline

 \Lambda(1810) &{\bf 23}\times 79=1817& \Sigma(1670) &{\bf 23}\times 73=1679\\ 
 & & &22\times 76=1672 \\  \hline

 \Lambda(1820) &{\bf 23}\times 79=1817& \Sigma(1690) &{\bf 23}\times 74=1702\\   
  &24\times 76=1824 & &22\times 77=1694 \\ \hline

 \Lambda(1830) &{\bf 23}\times 80=1840& \Sigma(1750) &{\bf 23}\times 76=1748\\                                              
&25\times 73=1825 &  & \\ \hline

 \Lambda(1890) &{\bf 29}\times 65=1885& \Sigma(1775) &{\bf 23}\times 77=1771\\ \hline

 \Lambda(2000) &{\bf 29}\times 69=2001& \Sigma(1840) &{\bf 23}\times 80=1840\\ \hline

 \Lambda(2020) &{\bf 29}\times 70=2030& \Sigma(1880) &{\bf 29}\times 65=1885\\ 
 &  &  &26\times 72=1872\\ \hline

 \Lambda(2100) &{\bf 29}\times 72=2088& \Sigma(1915) &{\bf 29}\times 66=1914\\ 
 & 30\times 70=2100&  & \\ \hline

 \Lambda(2110) &{\bf 29}\times 73=2117& \Sigma(1940) &{\bf 29}\times 67=1943\\ \hline

 \Lambda(2325) &{\bf 31}\times 75=2325& \Sigma(2000) &{\bf 29}\times 69=2001\\ \hline

 \Lambda(2350) &{\bf 31}\times 76=2356& \Sigma(2030) &{\bf 29}\times 70=2030\\ \hline

 \Lambda(2585) &{\bf 37}\times 70=2590& \Sigma(2070) &{\bf 29}\times 71=2059\\ \hline

 &  & \Sigma(2080) &{\bf 29}\times 72=2088 \\ \hline

  &  & \Sigma(2100) &{\bf 30}\times 70=2100 \\ \hline

  &  & \Sigma(2250) &{\bf 31}\times 73=2263 \\ 
 & & & 30\times 75=2250 \\ \hline

  & & \Sigma(2455) &{\bf 31}\times 79=2449 \\ 

  & &  &{\bf 37}\times 66=2442 \\ \hline

  & & \Sigma(2620) &{\bf 37}\times 71=2627\\ \hline

  & & \Sigma(3000) &{\bf 41}\times 73=2993\\ \hline

  & & \Sigma(3170) &{\bf 43}\times 74=3182\\ \hline

\end{array}
\end{displaymath}

\begin{displaymath}
\begin{array}{|c|c|c|c|} \hline

\Xi \mbox{baryon}& \mbox{Computed mass}

 &\Omega \mbox{baryon}& \mbox{Computed mass} 

\\ \hline

 \Xi(1317) &{\bf 19}\times 69=1311&  & \\ 
 \hline

 \Xi(1530) & 21\times 73=1533& \Omega(1672) &{\bf 23}\times 73=1679\\ 

 & {\bf 19}\times 81=1539&  & \\ \hline

 \Xi(1630) &{\bf 23}\times 71=1633& \Omega(2250) &{\bf 31}\times 73=2263\\  & & &30\times 75=2250 \\ \hline

 \Xi(1680) &{\bf 23}\times 73=1679& \Omega(2380)&{\bf 31}\times 77=2387\\ 
 &24\times 70=1680 & &33\times 72=2376 \\ \hline

 \Xi(1820) &{\bf 23}\times 79=1817& \Omega(2470) &{\bf 37}\times 67=2479\\ \hline

 \Xi(1940) &{\bf 29}\times 67=1943&  & \\ \hline

 \Xi(2030) &{\bf 29}\times 70=2030&  & \\ \hline

 \Xi(2120) &{\bf 29}\times 73=2117&  & \\ \hline

 \Xi(2250) &{\bf 31}\times 73=2263&  & \\ 
 &30\times 75=2250 & & \\ \hline

 \Xi(2370) &{\bf 31}\times 76=2356&  & \\
  &33\times 75=2375 & & \\ \hline

 \Xi(2500) &{\bf 31}\times 81=2511&  & \\

   & 32\times 78=2496&  & \\ \hline
\end{array}
\end{displaymath}

To conclude, we have given a computation of the mass spectrum of mesons and baryons. From the above tables we see that our computation can approximate the experimental masses of mesons and baryons. From this computation we show that there is a consecutive numbering of the mass spectrum of mesons and baryons and it is interesting that in this numbering many stable mesons 
and baryons are assigned with a prime number. In addition to the Regge theory and the constitutent quark model (CQM)\cite{Col}\cite{Lan}\cite{Clo} this computational method may provide a way to the study of the mass spectrum of mesons and baryons. We shall give a model of mesons and baryons from which this computation is based on. In this model each meson or baryon will be considered as a knot.

 \end{document}